\documentclass
[pra,superscriptaddress,showpacs,
twocolumn,
aps,floatfix,
  10pt]{revtex4-1}%
\usepackage{hyperref}
\usepackage[pdftex]{graphicx}
\usepackage{bm}
\usepackage{etoolbox}
\usepackage[usenames,dvipsnames]{color}
\usepackage{amsmath}
\usepackage{amsfonts}
\usepackage{amssymb}%
\newcommand{\bg}{ \begin{gather} }
\newcommand{\eg}{\end{gather}}
\newcommand{\be}{ \begin{equation} }
\newcommand{\ee}{\end{equation}}
\newcommand{\bea}{ \begin{eqnarray} }
\newcommand{\eea}{\end{eqnarray}}

\def \be {\begin{equation}}
\def \ee {\end{equation}}
\def \bea {\begin{align}}
\def \eea {\end{align}}
\def \p {\partial}
\def \BEA {\begin{eqnarray}}
\def \EEA {\end{eqnarray}}

\def \BC {\begin{cases}}
\def \EC {\end{cases}}

\usepackage{graphicx}
\usepackage{dcolumn}
\usepackage{bm}
\usepackage{amssymb}
\usepackage{amsmath}
\usepackage{verbatim}
\usepackage{color}
\usepackage{ulem}
\usepackage{epsf}
\usepackage{graphicx}

\begin{document}
\title{Asymmetry  of non-local dissipation: From drift-diffusion to hydrodynamics}

\author{K.\,S.~Tikhonov}
\affiliation{Institut f\"ur Nanotechnologie,  Karlsruhe Institute of Technology,
76021 Karlsruhe, Germany}
\affiliation{L.\,D.~Landau Institute for Theoretical Physics, 119334 Moscow, Russia}
\affiliation{Skolkovo Institute of Science and Technology, 143026 Skolkovo, Russia}
\author{I. V.~Gornyi}
\affiliation{Institut f\"ur Nanotechnologie,  Karlsruhe Institute of Technology,
76021 Karlsruhe, Germany}
\affiliation{A. F.~Ioffe Physico-Technical Institute,
194021 St.~Petersburg, Russia}
\affiliation{\mbox{Institut f\"ur Theorie der kondensierten Materie,  Karlsruhe Institute of
Technology, 76128 Karlsruhe, Germany}}

\author{ V. Yu.~Kachorovskii}
\affiliation{A. F.~Ioffe Physico-Technical Institute,
194021 St.~Petersburg, Russia}
%\affiliation{L.\,D.~Landau Institute for Theoretical Physics, 119334 Moscow, Russia}
\affiliation{Institut f\"ur Nanotechnologie,  Karlsruhe Institute of Technology,
76021 Karlsruhe, Germany}
\affiliation{CENTERA Laboratories, Institute of High Pressure Physics, Polish Academy of Sciences, 01-142 Warsaw, Poland }
\author{A. D.~Mirlin}
\affiliation{Institut f\"ur Nanotechnologie,  Karlsruhe Institute of Technology,
76021 Karlsruhe, Germany}
\affiliation{\mbox{Institut f\"ur Theorie der kondensierten Materie,  Karlsruhe Institute of
Technology, 76128 Karlsruhe, Germany}}
\affiliation{Petersburg Nuclear Physics Institute, 188300 St.Petersburg, Russia}
\affiliation{L.\,D.~Landau Institute for Theoretical Physics, 119334 Moscow, Russia}

\begin{abstract}
We study dissipation in inhomogeneous two-dimensional electron systems. We predict a relatively strong current-induced spatial asymmetry in the heating of the electron and phonon systems -- even if the inhomogeneity responsible for the electrical resistance is symmetric with respect to the current direction. We also show that the heat distributions in the hydrodynamic and impurity-dominated limits are essentially different. In particular, within a wide, experimentally relevant interval of driving fields, the dissipation profile in the hydrodynamic limit turns out to be asymmetric, and the characteristic spatial scale of the temperature distribution can be controlled by the driving field. By contrast, in the same range of parameters, impurity-dominated heating is almost symmetric, with the size of the dissipation region being independent of the field. This allows one to distinguish experimentally the hydrodynamic and impurity-dominated limits. Our results are consistent with recent experimental findings on transport and dissipation in narrow constrictions and quantum point contacts.
\end{abstract}
\maketitle

\section{Introduction}
\label{s1}

Electron transport involves two key ingredients: charge and energy transfer. Electrical resistance and heat dissipation, while always occurring back-to-back, typically rely on different mechanisms: elastic scattering off inhomogeneities and inelastic electron-phonon scattering, respectively. Understanding the underlying dynamics and the nature of dissipation is of fundamental importance and is also crucial for practical applications, in particular, in devices exploiting phase-coherent phenomena. Notably, resistance and dissipation in nanosystems can be dominated by spatially separated parts of the system. Such a ``heat-resistance separation'' is particularly prominent in ultraclean structures, as was discussed in detail in the seminal paper~\cite{rokni1995joule} for the case of a point contact. The derivation in Ref.~\cite{rokni1995joule} yielded two conceptually important results: (i) Joule heating is non-local and spatially separated from the contact (where the voltage drops); (ii) in the limit of small current, non-local heating is symmetric for symmetric contacts. The interpretation of the result (ii) was based on the assumption about the electron-hole symmetry at the Fermi level.

Impressive recent progress in nanoscale thermal measurements \cite{Thermo1,Thermo2,Thermo3,Thermo4,Thermo5,Thermo6,Thermo7,Thermo8,Thermo9,Thermo10,Thermo11,Thermo12}  has made it possible to test these statements with extremely high precision. In particular, a highly sensitive experimental method of thermal nanoimaging using a superconducting quantum interference device on a tip has been developed \cite{halbertal2016nanoscale}. This technique provides direct visualization of the dissipation mechanisms in quantum systems down to the spatial scale of a single impurity, with thermal sensitivity on the order of microkelvins.
This high-resolution thermography was employed to study dissipation in graphene~\cite{Halbertal2017}, where dissipation ring-shaped spots were observed in the bulk and on the edge of the samples, and  associated with individual atomic defects. This interpretation was supported by the theory of ``resonant supercollisions'' \cite{my-supercollisions,levitov-resonant}.  Although Ref.~\cite{Halbertal2017} did not address  the case of a point contact discussed in Ref.~\cite{rokni1995joule},  the reported results \cite{Halbertal2017}
clearly indicated the nonlocality of dissipation, in a full agreement with the general statement (i) of  Ref.~\cite{rokni1995joule}.  On the other hand, preliminary study \cite{Zeldov-unpublished} focused on the direct analysis of dissipation in symmetric point contacts  demonstrated that overheating of narrow constriction is asymmetric with respect to direction of the electric current. This observation should be contrasted to the statement (ii) of Ref.~ \cite{rokni1995joule} and thus requires further theoretical analysis.

Thermal nanoimaging experiments in ultraclean systems are also very interesting in view of recent discussion of signatures of hydrodynamical behavior in electrical and thermal transport at nanoscale (see Ref. \cite{Narozhny} and references therein). One of the key purposes of the current paper is to explore manifestations and hallmarks of the hydrodynamics in the the spatial character of dissipation.

Motivated by the recent experimental advances in thermal nanoimaging described above, we study in this paper the dissipation in a narrow constriction in a two-dimensional (2D) electron system.
We predict a relatively strong current-induced spatial asymmetry in the heating of the electron and phonon systems -- even if the inhomogeneity responsible for the electrical resistance is symmetric with respect to the current direction.
As we will show below, the spatial asymmetry of non-local dissipation can be explained within the framework of a kinetic equation taking into account electron-hole asymmetry in the vicinity of the Fermi level.
We will present calculations for both the hydrodynamic (HD) regime, which emerges when the electron-electron collisions dominate over other scattering mechanisms, and the impurity-dominated (ID) regime realized in dirty systems. While the hydrodynamic solution is rather straightforward, the calculation in the ID limit is more involved and requires specification of the electron-phonon collision integral. We use here a simplified model form of this integral which captures all physical properties of the problem and allows for an exact analytical solution.
We identify regions of parameters with different behavior of the dissipation profile and present analytical solutions for all of them. The control parameters are ratios of three characteristic length scales characterizing the size of the constriction, the current, and the electron scattering. We show that a relatively strong spatial asymmetry of dissipation arises generically when the current is not too weak. Furthermore, we demonstrate that the asymmetry of dissipation dramatically increases in the HD regime. Therefore, experimental observation of a very strong asymmetry, as in Ref.~\cite{Zeldov-unpublished}, represents an evidence of hydrodynamic type of transport.

%%%%%%%%%%%%
%%%%%%%%%%%%
\begin{figure}[ptb]
\centering
\includegraphics[width=0.48\textwidth]{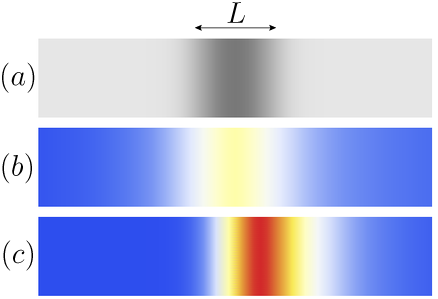}
\caption{(a) Inhomogeneous strip with a symmetric spatial dependence of the elastic scattering rate $1/\tau(x)$ (darker region corresponds to stronger scattering). Temperature profile in the strip in the ID (b) and HD (c) regime for $l_{\rm in}=l_{*}=4L$, where $L$ is characteristic size of inhomogeneity (see the text).}
\label{fig-1}
\end{figure}
%%%%%%%%%%%%%%%
%%%%%%%%%%%%%%%

\section{Model}
\label{Sec:Model}
We consider electron transport in a 2D system which consists of a narrow strip with an inhomogeneous distribution of the elastic scattering rate, see Figs.~\ref{fig-1}a. In this setup, the term ``constriction'' will be used for the region of enhanced elastic scattering (a macroscopic ``defect'' with increased resistance).
As we will see, dissipation in this model is qualitatively similar to that in a geometric constriction with homogeneous disorder. At the same time, the disorder-controlled constriction model allows one to simplify the solution by formally reducing the problem to a one-dimensional one.

We assume a parabolic dispersion for electrons characterized by mass $m$ and start from kinetic equation describing the distribution of electrons over velocity $\mathbf V$ in the electric field characterized by the force $\mathbf F$:
\be
\frac{\p f}{\p t} +\mathbf V \frac{\p f}{\p \mathbf r}
+ \frac{\mathbf F}{m} \frac{\p f}{ \p \mathbf V}=\widehat{\rm St } f.
\label{kinur}
\ee
Here,
$$\widehat{\rm St} = \widehat{\rm St}_\text{imp} +\widehat{\rm St}_\text{ph}
+\widehat{\rm St}_\text{ee}$$
is the collision integral including contributions from impurity, electron-phonon and electron-electron scattering, respectively.
We write the impurity collision integral in a standard form
\be
\widehat{\rm St}_\text{imp}f=\frac{ f_0 - f }{\tau},
\ee
where $\tau$ is the (coordinate-dependent) momentum relaxation time,
$f_0= \langle f\rangle$,
and $\langle \cdots \rangle$ stands for averaging over velocity directions.

In order to study electron-phonon heat balance, one needs to specify the electron-phonon collision integral.
The simplest model of this integral, which leads to relaxation to the Fermi distribution function with the lattice temperature $T_0$, reads
\be
\widehat{\rm St}_\text{ph}f
=\gamma \frac{\p  }{\p \epsilon}
\left\{\epsilon \left[f_0(1-f_0) + T_0\frac{\p f_0}{\p \epsilon}\right]\right\},
\label{St-ph}
\ee
where $\gamma$ is the electron-phonon scattering rate which is assumed throughout the paper to be energy-independent and small compared to the momentum relaxation rate:  $\gamma \tau \ll 1$. The collision integral (\ref{St-ph}) captures basic physics of the electron-phonon energy transfer and allows for exact analytical solutions.  It possesses the key properties of the electron-phonon collision integrals: it vanishes in equilibrium, conserves the total number of electrons, and does not transfer energy to $\epsilon<0$. The Fokker-Planck form of the collision integral (\ref{St-ph}) can be microscopically derived for the quasi-elastic scattering by acoustic phonons~\cite{perelbook}  for $T \gg T_{\rm BG},$ where $T_{\rm BG}$ is the Bloch-Gr\"uneisen temperature which determines   the maximum energy transferred from electron to acoustic phonons in a  collision process (in the absence of impurity-assisted  ``supercollisions'' \cite{Song,my-supercollisions,levitov-resonant}).  

As for the electron-electron collision integral $\widehat{\rm St}_\text{ee}$, we do not need an explicit expression for it and only use the fact that it preserves the total particle number, energy, and momentum. For simplicity, we characterize $\widehat{\rm St}_\text{ee}$ by a single electron-electron collision time $\tau_{\rm ee}$.

Since the system under the consideration is inhomogeneous along the $x$ direction, the electric field depends on the coordinate and  can be written as
\be
F(x)=F_0 + \delta F(x),
\ee
where $\delta F(x)$ is the inhomogeneity-induced correction which should be found self-consistently by solving the Poisson's equation. The calculations drastically simplify in the limit of infinitely strong screening, when  $\delta F$ can be found from the quasineutrality condition $N\approx N_0 =\text{const}$. In this paper, we will restrict ourselves to this limit only.

Below, we use different approaches depending on the relation between  momentum relaxation time $\tau$  and time of electron-electron collision, $\tau_{\rm ee}$.
For the case of fast electron-electron collisions ($ \tau_{\rm ee} \ll \tau$, we  use a hydrodynamic ansatz, while for slow collisions we neglect the electron-electron collision integral, assuming that the thermalization occurs solely because of the electron-phonon interaction.

The strength of overheating and the degree of dissipation asymmetry depend on the relation between characteristic lengths in the problem.
Specifically, one can conveniently introduce two length scales characterizing the energy transfer in the problem. The first is the diffusive length of inelastic scattering,
\be
l_* \sim \sqrt{D_0/\gamma},
\label{eq-lstar}
\ee
where $D_0$ is the diffusion coefficient in the absence of driving electric field.
The second is the drift inelastic length
\be
l_{\rm in} \propto v/\gamma,
\label{eq-lin}
\ee
which is proportional to the drift velocity $v$ governed by the electric field [see Eqs.~\eqref{G} and \eqref{lin} below].

\begin{figure}[ht]
\centering
\includegraphics[width=0.48\textwidth]{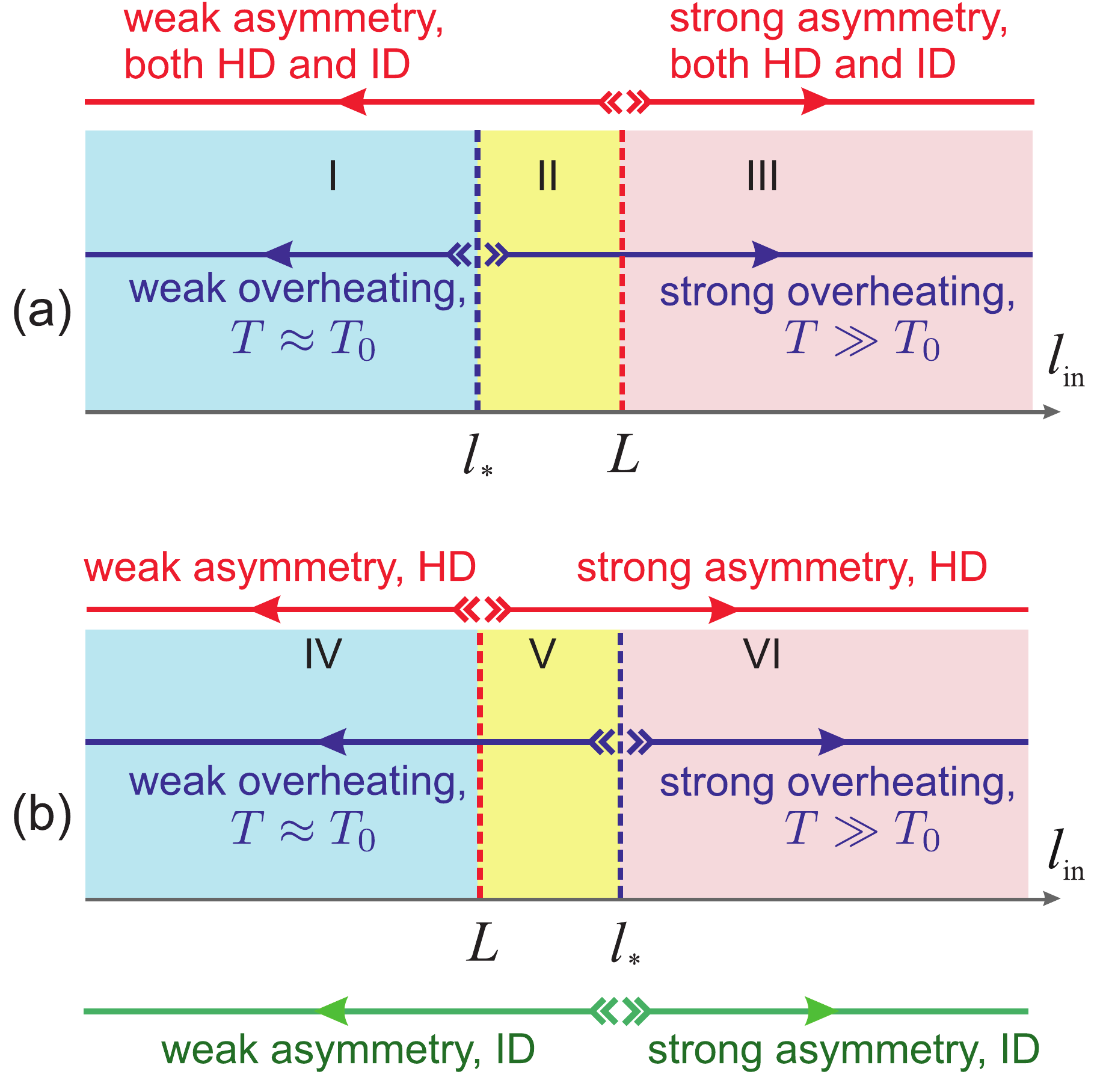}
 \caption{Schematics of dissipation regimes realized in the Boltzmann case for different relations between the field-independent lengths $l_*$ (diffusive electron-phonon length)
and $L$ (size of the constriction) and the drift-inelastic length $l_{\rm in}$ proportional to the driving electric field. Panel (a): $L>l_*$, regions I, II, and III are realized with increasing $F_0$. Panel (b): $L<l_*$, regions IV, V, and VI are realized with increasing $F_0$.}
\label{fig-2}
\end{figure}

In Fig.~\ref{fig-2}, for simplicity, we illustrate the heating regimes in the Boltzmann case, $ T_0\gg E_F$, where $E_F$ is the Fermi energy. In this case, $D_0  \sim T_0\tau/m$.   Physically,  the  lengths $l_*$ and $l_{\rm in}$ characterize inelastic scattering for weak and strong driving fields, respectively. One can introduce the ``true'' inelastic length $l_*(T)$ which in the Boltzmann case reduces to $l_*$ and $l_{\rm in}$ in the limiting cases:
\be l_*(T) \sim \sqrt{\frac{D(T)}{\gamma}}= \left\{  \begin{array}{c}
                                                      l_*,\quad  l_{\rm in } \ll l_* \\
                                                       l_{\rm in},\quad  l_{\rm in } \gg l_*
                                                    \end{array}
\right.,
\label{l*(T)}
\ee
where $D(T) \sim T \tau/m, $ and  $T$ is the temperature in the presence of the field [see Eq.~\eqref{T-F0} below]. 

Two panels of Fig.~\ref{fig-2}  correspond to the cases $l_* < L$ (Fig.~\ref{fig-2}a)  and  $l_* > L$ (Fig.~\ref{fig-2}b), where
$L$ is the size of the constriction. Further, the temperature distribution strongly depends on the relation between these two field-independent lengths $L$, $l_*$ and
the drift length $l_{\rm in}$.
In total, we have six different cases of ordering of the lengths $L$, $l_*$ and $l_{\rm in}$, which are labeled by roman numerals from I to VI in Fig.~\ref{fig-2}.

As we show below, the difference between the HD and ID regimes is particularly pronounced in the parameter region $L < l_{\rm in} <  l_*$ labeled V in Fig.~\ref{fig-2}b.
 In this case, the temperature distribution  in the ID regime is almost symmetric (with overheating  proportional to $F_0^2$ in accordance with earlier prediction of Ref.~\cite{rokni1995joule}) and has a small ($\propto F_0^3$) asymmetric correction. The spatial size of this distribution is on the order of $l_*$ and thus is field-independent. By contrast, the HD temperature distribution is strongly asymmetric with the characteristic scale on the order of $l_{\rm in}$ and, therefore, can be controlled by electric field.
This difference (see panels b and c of Fig.~\ref{fig-1}) can be experimentally resolved, giving a possibility to distinguish experimentally between hydrodynamic and drift-diffusion cases.

\section{Hydrodynamic regime}
\label{hydro}

\subsection{Hydrodynamic formalism}

In this section, we assume that
\be
\tau_{\rm ee} \ll \tau,
\label{cond-hydro}
\ee
(which also implies $\tau_{\rm ee} \ll \gamma^{-1}$)
so that fast electron-electron collisions drive the system into the HD regime,
in which the system is fully described by local values of drift velocity,
$\mathbf v(\mathbf r,t)$, temperature, $T(\mathbf r,t)$, and chemical potential
$\mu(\mathbf r,t)= N(\mathbf r,t)/\nu$.
(Here $N$ is the electron concentration and $\nu=\text{const}$ is the thermodynamic density of states.)  Although the derivation of hydrodynamic equations for these quantities is quite standard and can be found in textbooks,  we present this derivation in Appendix \ref{app:hyd} in order to make the presentation self-contained.

The hydrodynamic heat balance equation reads
\be
C\left[ \frac{\p T}{ \p t} +{\rm div} (\mathbf v T)\right]=N\left[\frac{mv^2}{\tau}-\gamma ~ (T-T_0) \right],
\label{dC-dt}
\ee
where $C$ is the heat capacitance of a 2D system given by
$C \approx {\pi^2 \nu T}/{3}$ for  $\nu T \ll N$  (Fermi distribution),
and $C \approx N$ for $\nu T \gg N$ (Boltzmann distribution).
In Eq.~\eqref{dC-dt} we neglected the second-derivative term with the heat conductivity, which is proportional to $\tau_{\rm ee}$ in the HD limit and is, therefore, small. We will discuss the role of this term in the end of the paper.

The temperature dependence of the electron-phonon term in Eq.~(\ref{dC-dt}) corresponds to the collision integral (\ref{St-ph}) with energy-independent $\gamma$. Indeed, when the Fermi function with $T\neq T_0$ is substituted in Eq.~(\ref{St-ph}), the result is proportional to $T-T_0$.
One can generalize Eq.~\eqref{dC-dt} for a more
general collision integral beyond the Fokker-Planck approximation   by replacing
\be
\gamma\: (T-T_0) \to\gamma \: \frac{T^k-T_0^k}{k\,T_0^{k-1}},
\label{dC-dt-k}
\ee
where the integer number $k$ depends on material and  the type of phonons (e.g., for graphene,   see Refs. \cite{Song,my-supercollisions}   and references therein). Assuming that the inhomogeneity leads to a small deviation, $\delta T(x)$,  of temperature from the value of $T$ at $|x| \to \infty$, linearization of collision integral yields $\gamma_T\delta T,$ where $$\gamma_T=\gamma (T/T_0)^{k-1}.$$

Let us make a short comment before solving Eq.~\eqref{dC-dt}. It was found
in Ref.~\cite{rokni1995joule} under the assumption of electron-hole symmetry, that the temperature distribution in the overheated system is a symmetric function with respect to the current direction. As we will show below, the breaking of the particle-hole symmetry gives rise to asymmetric temperature distribution, even when the deposited heat, described by the term ${mv^2}/{\tau}$, is a symmetric function of $x$. The particle-hole asymmetry reveals itself in Eq.~\eqref{dC-dt} through the term
${\rm div} (\mathbf v T)$. It is worth noting that this term is also responsible for nonzero thermopower.

Throughout the paper we will focus on calculation of the electron temperature distribution $T(x)$.  What is measured in experiment is the phonon temperature $T_0$ which also becomes position dependent because of the energy transfer between the electron and phonon subsystems: $T_0 \to T_{\rm ph}(x)$.
This dependence can be directly found from the heat balance equation for the phonon subsystem,
\be
-\kappa_{\rm ph} \Delta T_{\rm ph}
=\gamma \: (T-T_{\rm ph})-\gamma_{ 0}\: (T_{\rm ph}-T_{0}),
\ee
where $\kappa_{\rm ph}$ is the phonon heat conductivity, $T_{0}=\text{const}$ is the temperature of the substrate and $\gamma_{0}$ is the rate of the heat transfer between phonon subsystem and substrate. Typically, $\gamma_{0} \gg \gamma$,  and
$\kappa_{\rm ph} \ll \gamma_0 L^2$, so that $T_{\rm ph}$ is very close to the temperature of the substrate with a small correction which is fully expressed via the electron temperature:
\be
T_{\rm ph}(x)   \approx T_0+\frac{\gamma}{\gamma_0} T(x).
\label{Tph}
\ee

\subsection{Homogeneous heating}

In the stationary homogeneous case ($\p/\p t\to 0, \nabla\to 0$), we find from Eqs.~\eqref{dN-dt}, \eqref{dv-dt}, and \eqref{dC-dt}:
\be
N=N_0=\text{const},\quad \frac{\mathbf v}{\tau}=\frac{\mathbf F_0}{m},\quad \frac{mv^2}{\tau}=\gamma \:(T-T_0),
\label{hom}
\ee
yielding a homogeneous temperature of the electron system
\be
T=T_0+\frac{F_0^2 \tau}{m\gamma}
\label{T-F0}
\ee
which differs from the substrate temperature by a conventional
quadratic-in-field term.
The parameter controlling the overheating is
\be
\alpha =\left(1+\frac{m\gamma T_{0}}{F_0^{2}\tau }\right)^{-1},
\label{alpha}
\ee
so that
\be
T=\frac{T_{0}}{1-\alpha }= \frac{F_0^2\tau}{m\gamma  \alpha}.
\ee
For strong overheating, $1-\alpha \ll 1$, we get
$T\approx {F_0^2\tau}/{m\gamma}$ and initial temperature $T_0$ drops out from all final equations.
For a more general collision integral, we find by means of  Eq.~\eqref{dC-dt-k}:
\be
T=T_0\left(1+\frac{ k F_0^2 \tau}{T_0 m\gamma}\right)^{1/k}.
\label{T-F0-k}
\ee

\subsection{Dissipation profile around an inhomogeneity}

Next, we assume that $\tau$ depends on $x$, see Fig.\ref{fig-1}a, with a limiting value $\tau_\infty$ at $x \to \pm \infty $. This model can also mimick the geometrical constriction, see Appendix~\ref{app:constriction}. Below, we will demonstrate that even for symmetric dependence $\tau(x)=\tau(-x)$, the dependence of the temperature is asymmetric. Physically, this asymmetry arises from the electron-hole asymmetry at the Fermi level.
Therefore, the effect becomes particularly strong for the Boltzmann distribution ($T \gg \mu$)  for which the electron-hole asymmetry is maximal.

We focus on the simplest case of very short screening length, when quasineutrality of the  electron liquid  dictates its incompressibility. The corresponding criterion is most transparent for a gated system (see Appendix \ref{app:hyd}) characterized by the electrical capacitance $\mathcal C$.
The incompressible regime is effectively realized when the plasma-wave velocity
$s=\sqrt{{e^2N_0}/{m \mathcal C}}$
is sufficiently large: $s^2 \gg  T/m.$
In this regime, putting in Eq.~\eqref{dv-dt}
\be
N\approx N_0 =\text{const}
\ee and using then the current conservation,
\be
v=|\mathbf v| \approx \frac{F_0\tau_\infty}{m}=
\text{const},
\label{v-infty}
\ee
one simplifies Eq.~\eqref{dC-dt}:
\be
Cv\frac{dT}{dx}
=N_0 \left[ \frac{mv^2}{\tau (x)} -\gamma \:\frac{T^k-T_0^k}{ k T_0^{k-1}}\right],
\label{dt-dx}
\ee
where $C \approx  C(N_0,T)$.

The temperature of the electron gas at $|x| \to  \infty$
is given by Eq.~\eqref{T-F0-k} with the replacement $\tau \to \tau_{\infty}$.
For strong field, $T$ can be much larger than $T_0$. For weak inhomogeneities,
one can linearize collision integral around $T$.
Let us  assume that $\tau $  has a small $x$-dependent correction and introduce
the dimensionless function
\be
\xi(x)=\frac{\tau_\infty}{ \tau(x)} -1 =\int \frac{dq}{2\pi} e^{iqx} \xi_q,
\quad \xi(x) \ll 1,
\label{eta}
\ee
Then, Eq.~\eqref{dt-dx},  linearized with respect to $\xi$, becomes
\be
\frac{d \delta T}{dx} +\frac{\delta T}{l_{\rm in}}= G(x),
\label{dT-dx-Q}
\ee
where $\delta T =T(x)-T$,
\be
\label{gdef}
G(x)=\frac{ F_0  N_0  }{ C} \xi(x),
\ee
and
\be
l_{\rm in}=\frac{C}{N_0} \frac{v}{\gamma_T }
\label{G}
\ee
is the drift inelastic length with $v$  given by Eq.~\eqref{v-infty}.
The latter simplifies for the Boltzmann case (where $C\approx N_0$) and for a simplified collision integral \eqref{St-ph} for which $k=1$ and $\gamma_T=\gamma   $:
\be l_{\rm in}= \frac{v}{\gamma}.
\label{lin}
\ee
Solution of Eq.~\eqref{dT-dx-Q} reads
 \be
 \delta T(x) = \int \limits_{-\infty} ^\infty dx' K(x-x') G(x'),
  \label{dT(x)}
 \ee
 where
 \be
 K(x)=\int\frac{dq}{2\pi}\frac{e^{i q x}}{i q+ 1/l_{\rm in}}
 = \theta(x)e^{-x/l_{\rm in}}
 \label{Kx}
 \ee
is strongly asymmetric function of $x$.

\begin{figure}[ptb]
\centering
\includegraphics[width=0.40\textwidth]{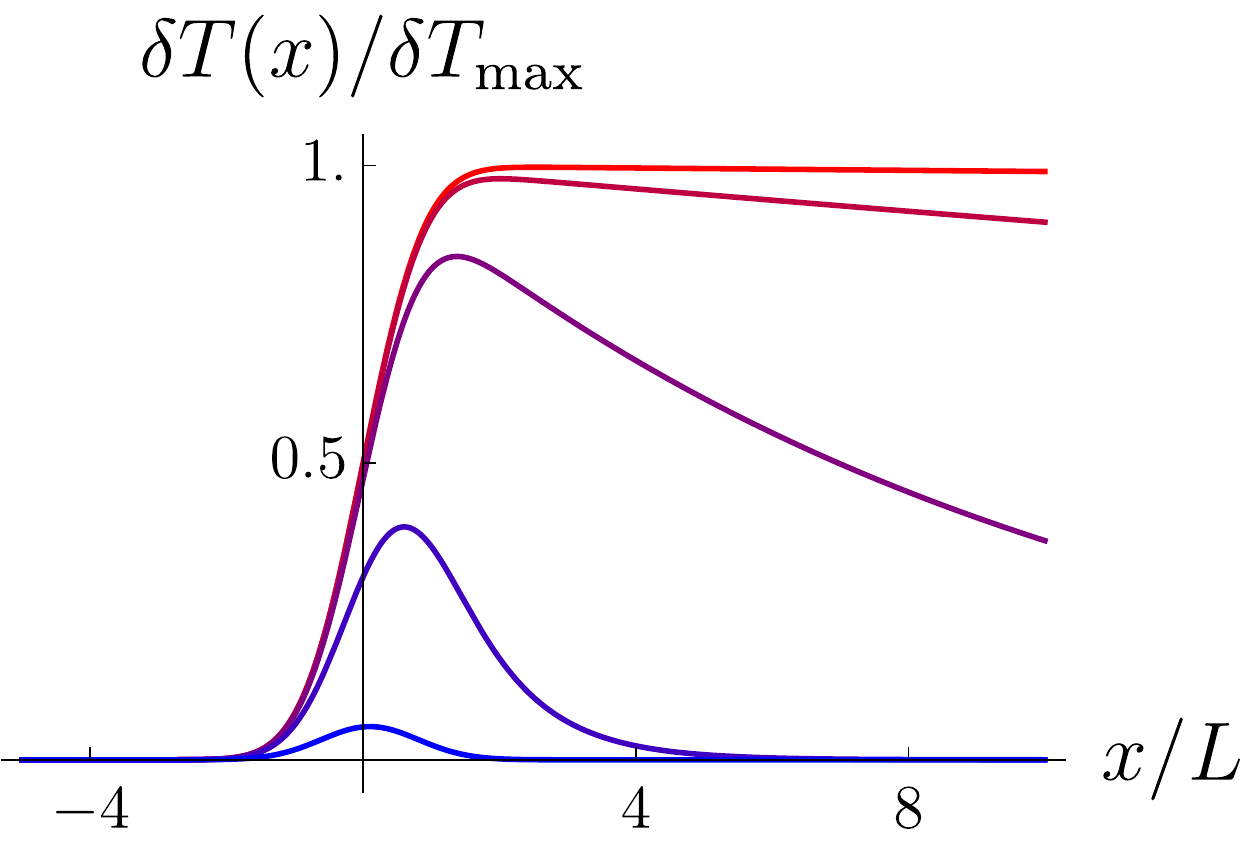}
\caption{ Spatial dependence of temperature in the HD regime for fixed $L$ and different $ l_{\rm in}$. From top to bottom:   $ L/l_{\rm in}=0.001,0.01,0.1,1,$ and $10.$   Dependence  $\tau(x)$ is modeled by the Gaussian shape with the width determined by $L$: ${1}/{\tau(x)}=\left({1}/{\tau_0}-{1}/{\tau_\infty}\right)\exp(- {x^2}/{L^2}) + {1}/{\tau_\infty}$ \cite{formula}.         }
\label{fig-3}
\end{figure}

The temperature profile   \eqref{dT(x)} is shown in Fig.~\ref{fig-3} for several values of   $L/l_{\rm in}$. For convenience, we assumed a Gaussian shape of the inhomogeneity $1/\tau(x)$ \cite{formula}.  It is seen from the figure that the asymmetry is very strong in the limit $L/l_{\rm in} \ll 1$ and becomes weak in the opposite limit. Below we analyze analytically these two limiting cases.

For weak coupling to the phonon system,
 \be
 L \ll l_{\rm in},
 \label{strong asymmetry}
 \ee
the shape of the function $\delta T(x)$ is strongly asymmetric: it decays for $x<0$ within the  distance  $L$  and  within  much longer distance $l_{\rm in}$ for $x>0$.
In the limit $l_{\rm in}= \infty$, a maximum asymmetry is reached and the
difference of temperatures at $x=\pm\infty$ tends to a finite value:
\be
 \delta T_{\rm max}= \frac{ F_0 N_0}{C}  \int \limits_{-\infty}^\infty \xi(x)dx,
 \label{dTmax}
\ee
Hence, the asymmetric part of the temperature distribution is proportional to the first power of the driving force $F_0$ and remains finite in the limit
$l_{\rm in} \to  \infty $ (for a fixed system size).
Fixing $l_{\rm in}$ and turning $L\to 0$, we find
 \be
 \delta T (x) \approx \delta T_{\rm max} K(x).
 \label{T(x)hydro}
 \ee

 %%%%%%%%%%%%%%%
%%%%%%%%%%%%%%%

In the opposite limit of fast electron-phonon collisions,
 $ L \gg l_{\rm in}$, the assymetry is weak. The temperature can be found by expanding $G(x')$ in the Taylor series near $x'=x$ in Eq.~\eqref{dT(x)}.
This yields
 \be
 \delta T(x) \approx  l_{\rm in} G(x) -  l_{\rm in}^2  \frac{dG(x) }{d x}.
 \label{T-expansion}
 \ee
The asymmetry is encoded in the second term which, for the Fokker-Planck collision integral \eqref{St-ph}, is proportional in the Boltzmann case to $(F_0^3/\gamma^2)  d\tau(x)/dx$, as follows from Eqs.~\eqref{lin}, \eqref{v-infty}, \eqref{gdef}, and \eqref{eta}. Hence, for weak electrical driving, the asymmetry arises in the third order with respect to electric field. With increasing field,
the asymmetry  becomes proportional to $F_0$ as follows from Eq.~\eqref{dTmax}.
This explains why this asymmetry is not captured by a conventional perturbative-in-field approach which accounts for heating effects within quadratic-in-field approximation \cite{rokni1995joule} (see a more detailed discussion in Sec.~\ref{comp}).

It is instructive to analyze how increasing the chemical potential (and thus decreasing degree of the electron-hole asymmetry) affects the dissipation regime.
For large $\mu$, we have
$l_{\rm in} \propto C/N \sim  T/\mu$. Therefore, for sufficiently large chemical potential, $l_{\rm in}$ becomes smaller than $L$ and we drive the system into regime of weak asymmetry described by Eq.~\eqref{T-expansion}.

To conclude this Section, we would like to stress that within the hydrodynamic  picture, inhomogeneity of overheating is only controlled by the relation between the size of the resistance inhomogeneity $L$ and drift inelastic length $l_{\rm in}$, as illustrated in Fig.~\ref{fig-2}. Parameter regions  I, II, and IV correspond to weak asymmetry of dissipation, while in the regions III, V, and VI the asymmetry is relatively  strong.

\section{Impurity-dominated regime}

\subsection{Kinetic equation formalism and general solution}
\label{kin-eq-formalism}

In the previous Section, we discussed the hydrodynamic limit, assuming that the momentum-conserving electron-electron collision time is the shortest scattering time.
Let us now consider the opposite case of dominating impurity scattering:
\be
1/\tau \gg \gamma \gg 1/\tau_{\rm ee},
\ee
when the electron-electron collision integral can be neglected.
For simplicity, we will restrict ourselves to the non-degenerate (Boltzmann) regime
and assume that the momentum-relaxation time is independent of energy.
We will also rely on the model form of the collision integral Eq.~\eqref{St-ph}.
Such a model allows for exact analytical solution.

We search for solution to Eq.~\eqref{kinur} in the standard form (see, e.g., Ref.~\cite{rokni1995joule})
\be
f=f_0(\mathbf r,\epsilon) + f_1(\mathbf r,\epsilon)e^{i\theta}
+ f_{-1}(\mathbf r,\epsilon)e^{-i\theta},
\label{f0f1}
\ee
where $\theta$ is the angle of velocity $\mathbf V$ and $\epsilon=m V^2/2$ is the particle energy. The neglect of higher angular harmonics $f_n$ with $|n|>1$ is justified provided that $\tau$ is the shortest time scale, such that elastic mean free path is shorter than both  inhomogeneity size $L$ and inelastic length $l_*(T)$ given by Eq.~\eqref{l*(T)}.

Substituting Eq.~\eqref{f0f1} into Eq.~\eqref{kinur}, projecting thus obtained equation onto $0$ and  $\pm 1$ angular harmonics, we get a closed set of equations for $f_0,f_1$ and $f_{-1}.$  Next,  expressing $f_{\pm 1}$ through $f_0$, we derive a closed equation for the isotropic part of the distribution function
\be
-\left(\frac{\p}{\p x}+F \frac{\p}{\p \epsilon }\right) D(\varepsilon)\left(\frac{\p}{\p x}+F \frac{\p}{\p \epsilon }\right)f_0= \widehat{\rm St}_{\rm ph}\, f_0,
\label{kinur for f0}
\ee
where $D(\epsilon)=\epsilon \tau/m$ is the energy-dependent diffusion coefficient.

For the homogeneous system $\tau(x)\equiv\tau$ one has
 \be
- F_0^2
 \frac{\p}{\p \epsilon } D(\varepsilon) \frac{\p}{\p \epsilon }f_0(\epsilon) = \widehat{\rm St}_{\rm ph}~ f_0.
\label{homogeneous}
\ee
For the model collision integral given  by Eq.~\eqref{St-ph}, the solution of Eq.~\eqref{homogeneous} gives the Boltzmann distribution
$$f_T= \nu N_0 \exp (-\epsilon/T)$$
with the temperature given by Eq.~\eqref{T-F0}.
For an inhomogenous system, we search for a solution to Eq.~\eqref{kinur for f0} by expanding $f_0$ in $\xi$ (see Eq.~\ref{eta}).
To this end, we write
\be
f_0=f_{T}+\delta f,
\label{deltaf}
\ee
where $\delta f \propto \xi$ is a small inhomogeneity-induced correction.
We linearize Eq.~\eqref{kinur for f0} with respect to $\delta f$, $\xi (x)$ and
\be
\lambda(x)=\delta F(x)/F_0
\label{lambda}
\ee
where
\be
\delta F (x)   = - \frac{e^2}{\mathcal C}\nabla \delta N
\label{dF}
\ee
is the electrostatic force induced by the density variation $\delta N =\nu \int d\epsilon \delta f $ (see Appendix \ref{app:hyd}).

The  kinetic equation linearized with respect to $\delta f$  acquires the  form
\be
\hat L \delta f = \mathcal { S}, \label{kin-lin}
\ee
where $\hat L$ is a linear operator and $\mathcal{S}$ is an energy-dependent  source. Exact expressions for $\hat L$  and $\mathcal{S}$  are given in Appendix \ref{app:int}.
Interestingly,  the operator $\hat L$ is non-Hermitian but has a discrete spectrum, which stems from the requirement that the distribution function should be finite both at
zero energy (one of the solutions is logarithmically divergent at $\epsilon \to 0$) and
at $ \epsilon =\infty $ (one of the solutions grows exponentially  at $\epsilon \to \infty$).

The explicit solution of Eq.~\eqref{kin-lin}   is presented in terms of the eigenmode expansion in Appendix \ref{app:int}.
The final result for the temperature distribution
\be
\delta T(x)=\int_{0}^{\infty }\left(\frac{\epsilon}{T} -1\right)
\delta f\left(x,\epsilon \right) d \epsilon
\ee
can be written in the form analogous to Eqs.~\eqref{dT(x)} and \eqref{gdef},
with the Fourier transform of the kernel given by Eq.~(\ref{KQ}) in Appendix \ref{app:int}. Let us now discuss various limiting cases of this kernel. To this end, we introduce the field-independent length
\be
 l_*=\sqrt{\frac{T_0 \tau_{\infty}}{m\gamma}},
 \label{l*}
 \ee
 which has a physical meaning of the diffusive energy relaxation length in weak fields when overheating can be neglected, i.e.
\be
\alpha \approx \frac{F^2\tau_\infty}{m \gamma T_0} \ll 1 \qquad \textrm{ and } \qquad T \approx T_0.
\ee
The length  $l_*$ was denoted $\tilde l$ in Ref.~\cite{rokni1995joule}.
For small $\alpha$,  we have $l_{\rm in}=\sqrt{\alpha} l_*$. Further consideration depends on the relation between $l_*$ and $L$, as shown in Figs.~\ref{fig-2}(a) and (b) for $L>l_*$ and $L<l_*$, respectively. Let us now discuss possible limiting cases.

\subsection{Large defect size}

At large $L$, the size of the macroscopic defect is the largest scale: $L\gg l_{\rm in}, l_*$.
This situation corresponds to the regions I and II in Fig.~\ref{fig-2}(a). In this case dissipation is almost local, the asymmetry is weak and $\delta T(x)$ is determined by local Joule heat with small non-local corrections. Technically, analytical expression for $\delta T$ can be found by expanding the Fourier-transformed dissipation kernel (\ref{KQ})
into series over the wave-vector. In order to find this expansion up to the third order in gradients $\partial_x$, it is enough to cut the sums in Eq. (\ref{KQ}) at $n=2$ and $m=1$ and replace in these sums $Q$ with $- i \partial_x.$ As a result, we find
\begin{align}
\delta T (x) \approx F_0 l_{\rm in}  \hat{\mathcal{K}}_0 \xi(x)
\label{dT-imp-F}
\end{align}
with
\be
\hat{\mathcal{K}}_0 =
1 -l_{\rm in}\partial_x +  \left( 1 + \frac{2}{\alpha}\right)l_{\rm in}^2\partial_x^2
- \left(1+\frac{8}{\alpha}\right) l_{\rm in}^3\partial_x^3
\label{K0}
\ee
and $\xi$ given by Eq.~\eqref{eta}.

\begin{figure}[ptb]
\centering
\includegraphics[width=0.48\textwidth]{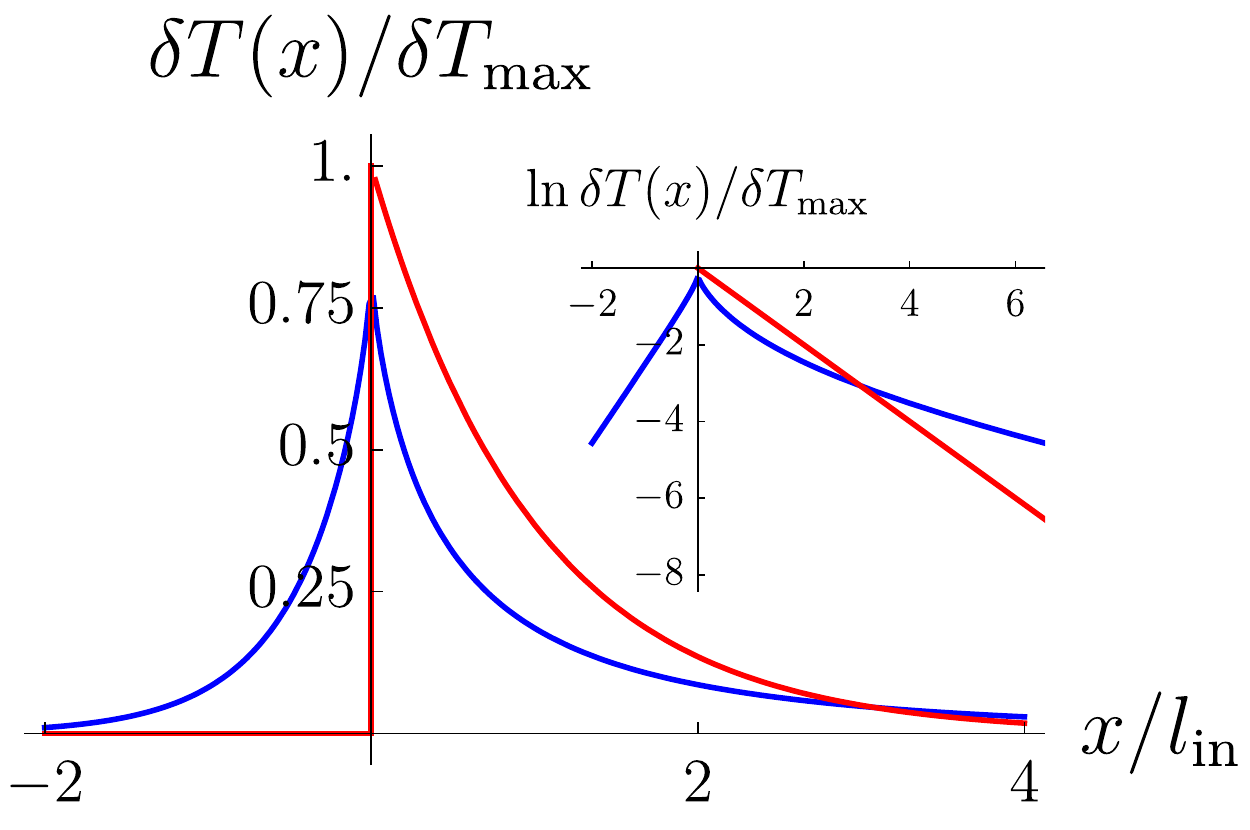}
 \caption{ Spatial dependence of temperature  in the hydrodynamic (red curve) and impurity-dominated (blue curve) regimes for the case $l_{\rm in} \gg L, \: l_*$
 (regions III and VI in Fig.~\ref{fig-2}).      }
\label{fig-4}
\end{figure}

\subsection{Large field}

At large $F$, the field-related scale $l_{\rm in}$ is the largest one: $l_{\rm in} \gg L, l_*$. This situation corresponds to the regions III and VI in Fig.~\ref{fig-2}. The overheating is strong in this case ($\alpha=1$) and $T\gg T_0$. The defect can be treated as point-like and, as a result,
$\delta T(x)/T $ is a universal function of $x/l_{\rm in}$. This function is plotted in Fig.~\ref{fig-4} by the blue line. It is interesting to note that $\int \delta T(x)dx$, as well as $\int x \delta T(x)dx$ are the same for the hydrodynamic and impurity-dominated regimes.

\subsection{Small field}

\begin{figure}[ptb]
\centering
\includegraphics[width=0.48\textwidth]{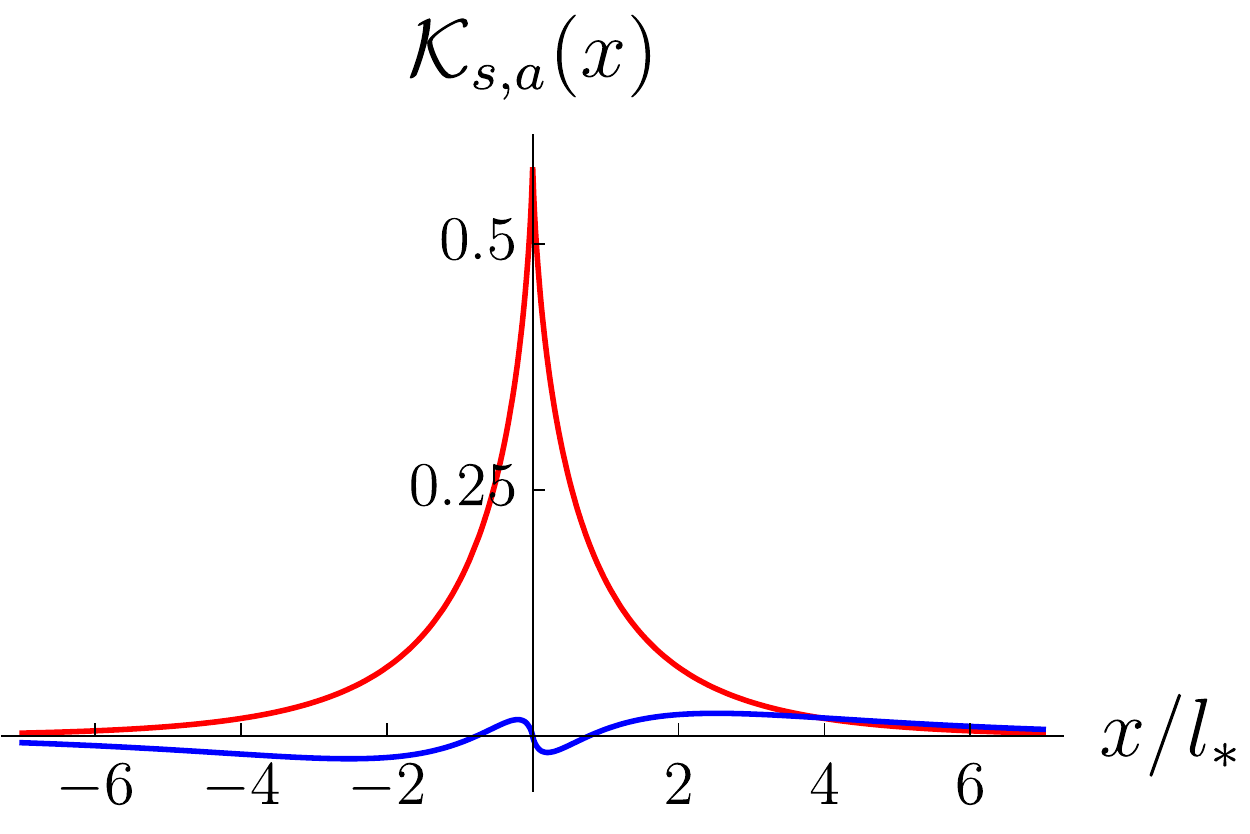}
 \caption{ Spatial dependence of
kernels $K_s$ (red curve) and $K_a$ (blue curve), Eq.~\eqref{Kas}.     }
\label{fig-5}
\end{figure}

At small $F$, when $l_{\rm in}\ll l_*$, the dissipation kernel $\hat{\mathcal{K}}$ relating $\delta T(x)$ and $\xi(x)$
can be expanded in $F$. In this case, overheating is small and characteristic inelastic length is given by $l_*$ [see Eq.~\eqref{l*(T)}]. This situation corresponds to the regions I, IV, and V in Fig.~\ref{fig-2}. Keeping the two leading terms in the expansion over $F$, we may write
\be
\delta T(x) =  F_0 l_{\rm in}\left[\hat{\mathcal{K}}_s + \frac{F l_*}{T} \hat{\mathcal{K}}_a\right]\xi(x)
\label{dt-imp-weak-K},
\ee
where $\hat{\mathcal{K}}_{s,a}$ are non-local integral operators with spatial scale $l_*$:
\be \hat{\mathcal{K}}_{s,a}\,\xi (x) = l_*^{-1}\int K_{s,a}\left(\frac{x-x'}{l_*}\right)\xi(x')dx'.
\label{Kas}
\ee
Kernels $K_{s}$ and $K_{a}$ are even/odd functions with respect to their argument, respectively.  Evaluation of these kernels requires calculation of the sums in Eq. (\ref{KQ}). Although they can be explicitly evaluated in terms of hypergeometric functions, the result is too cumbersome to be presented here. Instead, we plot these kernels in Fig.~\ref{fig-5}.

Assuming further that $L\ll l_*$, we find for the regions IV and V of Fig.~\ref{fig-2}:
\be
\delta T(x) = \frac{l_{\rm in}}{l_*} \delta T_{\rm max}  \left[K_{s} \left(\frac{x}{l_*}\right) + \frac{l_{\rm in}}{l_*} K_{a} \left(\frac{x}{l_*}\right) \right]
\label{dt-imp-weak}
\ee
with
\be
\label{dtmax}
\delta T_{\rm max} =F_0\int dx\, \xi(x).
\ee

The symmetric term is proportional to $F_0 {l_{\rm in}}/{l_*}\propto F^2$ and gives non-local symmetric overheating.  An analogous contribution to $\delta T(x)$ was found in Ref.~\cite{rokni1995joule} in the relaxation time approximation for  the electron-phonon collision integral.  Non-locality effects are controlled by the diffusive inelastic length $l_*$, in accordance with results of Ref.~\cite{rokni1995joule}. The asymmetric  term in Eq.~\eqref{dt-imp-weak} is proportional to $F_0^3$. It gives correction which is small in $l_{\rm in}/l_*$ in this regime,  and, therefore, leads to a weak asymmetry of the temperature distribution. This term is beyond the $F_0^2-$approximation used in Ref.~\cite{rokni1995joule}.

\subsection{Overlap of ``large-defect''  and ``small-field'' regimes}

Equations (\ref{dT-imp-F}) and (\ref{dt-imp-weak-K}) have a certain overlap in validity: region I. In particular, considering Eq. (\ref{dT-imp-F}) in the limit of small overheating $\alpha\to 0$ and considering Eq. (\ref{dt-imp-weak-K}) in the local limit $l_*\ll L$, when
$$
\hat{\mathcal{K}}_s=1+2 l_*^2\partial_x^2, \qquad \hat{\mathcal{K}}_a=-l_*\partial_x-8l_*^3\partial_x^3,
$$
we arrive to the same expression:
    \be
  \delta T (x)\! \approx\! F_0 l_{\rm in}\left[\left(1 + 2l_*^2 \partial_x^2 \right)
  \!-l_{\rm in} \! \left( \partial_x +8l_*^2 \partial_x^3\right)\right]\xi(x).
  \label{dt-im-FF}
  \ee

%In the next section we compare thus obtained temperature distribution with the %hydrodynamical solution

\section{Comparison of hydrodynamic and impurity-dominated regimes}
\label{comp}

Let us now compare the results obtained in the HD and ID limits. In the large-field regime corresponding to regions III and VI in Fig.~\ref{fig-2}, the hydrodynamic temperature distribution looks rather simple.
Indeed, in  this case, the function $\xi(x)$ is sharply peaked as compared to the  inelastic length $l_{\rm in}$ and, as a result [see Eq.~\eqref{Kx}],
\be \delta T= \delta T_{\rm max}\theta(x)\exp(-x/l_{\rm in}),
\label{HD-uni}
\ee
with $\delta T_{\rm max}$ given by Eq. (\ref{dtmax}) for the Boltzmann case
($C=N_0$).
We plot this dependence  in Fig.~\ref{fig-4} together with the corresponding dependence obtained in the ID strongly overheated regime ($\alpha=1 $).
We see that the hydrodynamical function is  much more asymmetric.
One of the physical reasons for  this difference is that electron-electron
collisions suppress heat conductivity, which turns out to be
proportional to $\tau_{\rm ee}$ and turns to zero in the purely hydrodynamical limit
of ideal fluid, $\tau_{\rm ee} \to 0$.

Let us now discuss what happens if we take into account small corrections with respect to $\tau_{\rm ee}$ allowing for a finite heat conductivity, $\kappa_0 \neq 0$, of the electron fluid. This modifies Eq.~\eqref{dT-dx-Q} as follows:
\be
\frac{d\delta T}{dx} - \rho \frac{d^2\delta T}{dx^2}= G(x) - \frac{ \delta T}{l_{\rm in}},
\label{dT-dx-Q-kappa}
\ee
 where
 \be
 \label{rhodef}
 \rho= \frac{\kappa_0}{Cv}
 \ee
 Solution of Eq.~\eqref{dT-dx-Q-kappa} is given by Eq.~\eqref{dT(x)}
 with
\BEA
K(x)&=&\int \limits_{-\infty}^{\infty}\frac{dq}{2\pi} \frac{e^{i q x}}{iq+ 1/l_{\rm in} +\rho q^2}
 \nonumber
 \\
 &=&\int \limits_0^\infty \frac{dt}{2\sqrt{\pi \rho t}}\exp\left[ -\frac{t}{l_{\rm in}}  -\frac{(x-t)^2}{4 t \rho} \right].
 \label{KK}
 \EEA
Analyzing Eq.~\eqref{KK}, we find that the heat conductivity does not affect the shape of the temperature distribution and can, therefore, be fully neglected provided that $$\rho \ll L,$$
or, equivalently
${CvL}/{\kappa_0} \gg 1.$
For  $ L \ll \rho  \ll   l_{\rm in}  $  the sharp jump in the temperature distribution at $x=0$  (see Fig.~\ref{fig-3}) gets broadened but the distribution is still asymmetric. Only for $\rho \gg l_{\rm in}$ the asymmetry starts to decrease.

For $l_* \ll L$ and  low field   $l_{\rm in}\ll L $ [regions I and II in Fig.~\ref{fig-2}(a)], the temperature correction  can also be found by expanding  the Fourier transform of the hydrodynamic kernel $K(x)$ [see Eqs.~\eqref{dT(x)} and \eqref{Kx}] into series over $q$. Then, we obtain:
\be
\delta T^{\rm hydro} (x) \approx F_0 l_{\rm in}\left[ 1 -l_{\rm in}\partial_x + l_{\rm in}^2\partial_x^2 - l_{\rm in}^3\partial_x^3\right]\xi.
\ee
Comparing this formula  with Eq.~\eqref{dT-imp-F}, we conclude that the hydrodynamic temperature distribution, in contrast to impurity-dominated one, is not sensitive to the overheating parameter.  This happens because in the hydrodynamic regime diffusive heat transfer is suppressed  and  the only relevant scale for energy relaxation  is the drift energy relaxation length $l_{\rm in}.$

The difference between HD and ID heating is most pronounced in region V in Fig.~\ref{fig-2}(b).  Here, the impurity-dominated overheating, described by the first term in Eq.~\eqref{dt-imp-weak}, is almost symmetric (with weak anisotropic correction $\propto K_{\rm as}$), and is dimensionless function of $x/l_*,$
while hydrodynamic overheating is strongly asymmetric and is given by a dimensionless function of $x/l_{\rm in}$ [see Eq.~\eqref{HD-uni}]. It is worth noting that such two dependencies can be easily distinguished experimentally because $l_{\rm in}$ is field-dependent in contrast to $l_*$.

\begin{figure}[ptb]
\centering
\includegraphics[width=0.40\textwidth]{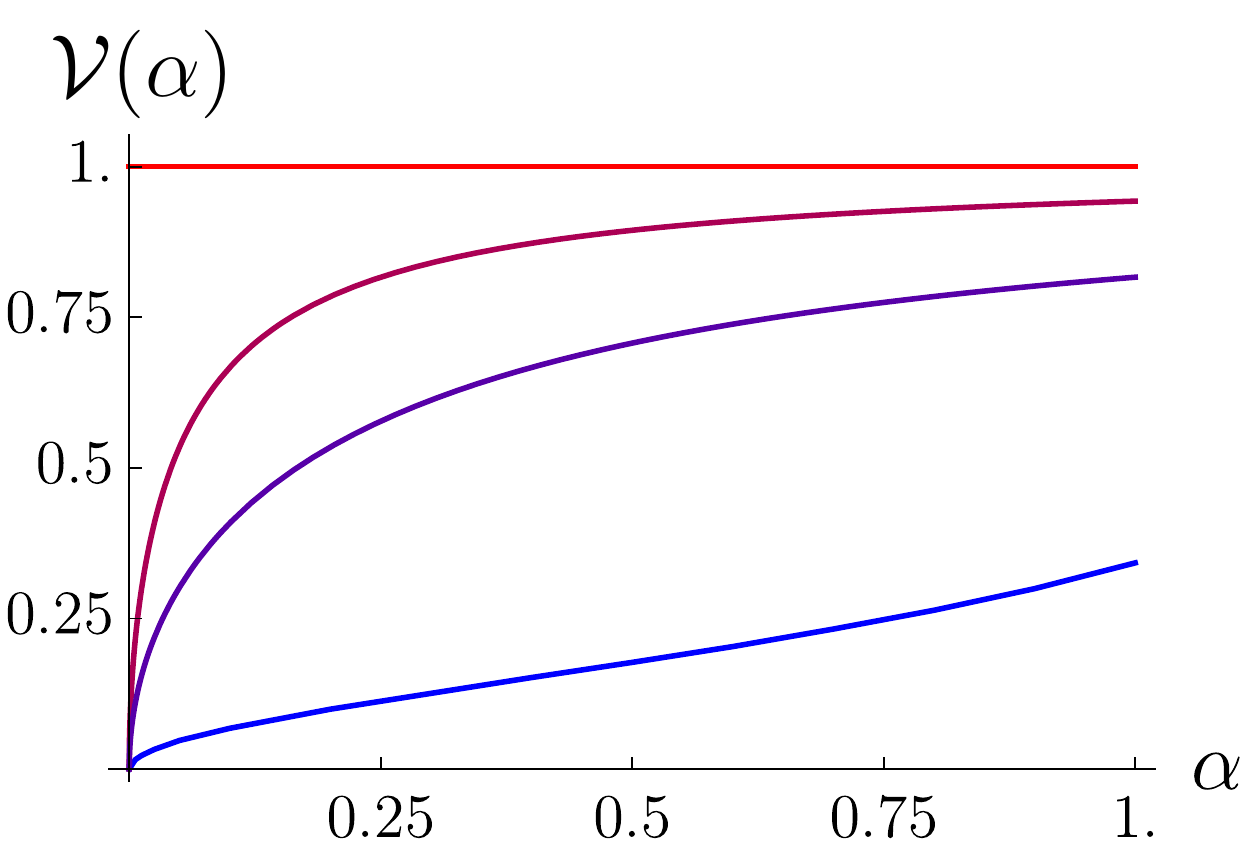}
\caption{Visibility of the asymmetry, $\mathcal{V}$ for a small inhomogeneity [$L\ll l_{\rm in},$  regions III, V and VI in Fig.~\ref{fig-2} (b)] as a function of the parameter $\alpha$ defined in Eq.~(\ref{l-star-general}). For the electron-phonon collision integral as in Eq.~(\ref{dC-dt}), i.e., with $k=1$, the parameter $\alpha$ is given explicitly by Eq.~(\ref{alpha}). The limits $\alpha \to 0$ and $\alpha\to 1$ correspond then to weak and strong field, respectively.
Upper line (red) --  hydrodynamic regime, $\tau_{\rm ee}/\tau \to 0$; visibility is maximal (equal to unity).  Lower line (blue) -- impurity-dominated regime, $\tau_{\rm ee}/\tau \to \infty$, calculated by using the results of Sec.~\ref{kin-eq-formalism} and Appendix \ref{app:int}.
Intermediate curves illustrate the crossover between the hydrodynamic and impurity-dominated limits, as described by Eq. (\ref{xV});  they correspond to $\tau_{\rm ee}/\tau=0.125$ and 0.5.}
\label{fig-6}
\end{figure}

To characterize the degree of asymmetry of the dissipation, we define the corresponding dimensionless visibility
$$
\mathcal{V}=\frac{\int_0^{\infty}\delta T(x)dx-\int_{-\infty}^0\delta T(x)dx}{\int_{-\infty}^{\infty}\delta T(x)dx}.
$$
In Fig. \ref{fig-6} we show the crossover in the visibility from the hydrodynamic to the impurity-dominated regime for the case of a small defect,
$L \ll (l_*,l_{\rm in})$.
In the ideal hydrodynamic limit, $\tau_{ee} \to 0$ ($\rho=0$), the visibility is   simply given by unity. For the hydrodynamics with a finite heat conductivity, the analytical expression for visibility is given by
\be
\mathcal{V}=\left(1 +\frac{4\rho}{l_{\rm in}}\right)^{-1/2}
\ee
In order to evaluate $\rho$ defined in Eq. (\ref{rhodef}),
we estimate the heat conductivity $\kappa_0$ in the HD regime as
$\kappa_0\sim(\tau_{\rm ee}/\tau) C D$. It follows that
$$\rho=\frac{T}{T_0}\left(\tau_{\rm ee}/\tau\right)l_{*}^2/l_{\rm in}.$$
Using now
\be
l_{*}/l_{\rm in}=\sqrt{{1}/{\alpha}-1},
\label{l-star-general}
\ee
where $\alpha$ is a dimensionless parameter defined in Eq.~(\ref{alpha}), we find
\be
\label{xV}
\mathcal{V}= \frac{\sqrt \alpha}{ \sqrt{\alpha +{4\tau_{\rm ee}}/{\tau}}}.
\ee

%%%%%%%%%%%%
%%%%%%%%%%%%
%%%%%%%%%%%%%%%
%%%%%%%%%%%%%%%

In the impurity-dominated regime, the visibility is calculated
by using the results for the spatial profile of temperature in Sec.~\ref{kin-eq-formalism} and Appendix \ref{app:int}. The corresponding asymptotic behavior of the visibility in the case of weak field (small $\alpha$) can be found
from Eq.~(\ref{dt-imp-weak}).   The result is  expressed in terms of the odd and even kernels (Fig. \ref{fig-5}) as
\be
\mathcal{V} = \sqrt{\alpha}\ \dfrac{\int_0^{\infty}K_a(x)dx}{\int_0^{\infty}K_s(x)dx} \approx 0.21\sqrt{\alpha}.
\ee

Finally, we remind the reader that the formula (\ref{alpha}) for the parameter $\alpha$ was derived for a particular form of the electron-phonon collision integral, as in Eq.~(\ref{dC-dt}), which corresponds to $k=1$. The results derived above  have, however, a general validity when expressed in terms of the relevant length scales. For a general collision integral, Eq.~\eqref{dC-dt-k}, we find from  Eqs.~\eqref{dT(x)} and \eqref{G}
\be
l_{\rm in}= \frac{C F_0\tau }{\gamma N_0 m } \frac{1}{\left(1+\frac{kF_0^2 \tau}{T_0 m \gamma} \right)^{(k-1)/k}}.
 \label{maximal}
 \ee
 The results shown in Fig.~\ref{fig-6} remain valid, with $\alpha$ defined now by Eq.~(\ref{l-star-general}) and with $l_{\rm in}$ from Eq.~(\ref{maximal}). Note that for $k>2$ the length $l_{\rm in}$ becomes a non-monotonous function of the electric field.

\section{Summary}

To summarize,  we have investigated dissipation in a narrow constriction in a two-dimensional electron  system.  Our main prediction is a rather strong current-induced asymmetry  in the heating of the electron and phonon systems, which is different in hydrodynamical and impurity-dominated regimes.  The spatial profile of the dissipation  in the hydrodynamic regime turns out to exhibit a particularly strong asymmetry, as illustrated in Fig.~\ref{fig-1}.  The corresponding spatial scale of the temperature distribution, $l_{\rm in}$,  can be controlled by the driving field.  By contrast,  the asymmetry of impurity-dominated heating is moderate, and the spatial scale of corresponding temperature distribution, $l_{*}$, does not depend on the field.   The degree of the asymmetry is  controlled by the parameter $\alpha$ that depends on the strength of the applied electric field, see Fig.~\ref{fig-6}. Our results  are consistent with recent experimental findings on  dissipation in narrow constrictions and quantum point contacts.

As further developments of our study, it would be worth considering other geometries (including point contacts), the effects of magnetic field, as well as effects of viscosity and boundary scattering in the hydrodynamic regime.

\section{Acknowledgments}

We are grateful to E. Zeldov for motivating discussions and sharing the unpublished experimental data with us. We also acknowledge collaboration with K. Dapper at the early stage of the work. We thank G. Zhang for carefully reading the manuscript and useful comments. This work is supported by the program 0033-2019-0002 by the Ministry of Science and Higher Education of Russia, by  the FLAGERA JTC2017 Project GRANSPORT through the DFG Grant
No. GO 1405/5, by RFBR (Grant No. 17-02-00217), by
Foundation for the Advancement of Theoretical Physics and Mathematics  ``BASIS'', and by the Foundation for Polish Science through the grant MAB/2018/9  for CENTERA. KT acknowledges support by Alexander von Humboldt Foundation.

\appendix

\section{Derivation of hydrodynamic equations}
\label{app:hyd}

In this Appendix, we provide a derivation of the hydrodynamic equations used in Sec.~\ref{hydro}.
We search for solution in the form of hydrodynamic ansatz
\be
f(\mathbf r,\mathbf V)=\dfrac{1}{\exp\left\{\dfrac{m\left[\mathbf V- \mathbf v(\mathbf r,t)\right]^2/2-\mu(\mathbf r,t)}{T(\mathbf r,t)} \right\}+1},
\label{dist-function}
\ee
where $\mathbf v(\mathbf r,t),\ T(\mathbf r,t)$, and $\mu(\mathbf r,t)= N(\mathbf r,t)/\nu, $ are, respectively,  local values of drift velocity, temperature and chemical potential, $N$ is the electron concentration and $\nu=\text{const}$ is the thermodynamic density of states.
Multiplying Eq.~\eqref{kinur} by ``$1$'', ``$\mathbf V$', and``$\epsilon=mV^2/2$'', integrating over $\mathbf V$ and using ansatz \eqref{dist-function},  after some algebra, we get the following set of equations:
\BEA
&&\!\!\frac{\p N}{\p t}+{\rm div}\left(\mathbf v N \right)=0,
\label{dN-dt}
\\
&&\!\!\frac{\p \mathbf v}{\p t}  +(\mathbf v \nabla) \mathbf v + \frac{\mathbf v}{\tau}=\frac{1}{m}\left( \mathbf{F}_0- \frac{e^2 \nabla N}{\cal C}  - \frac{1}{N} \nabla W \right),
\label{dv-dt}
\\
&&
\!\!\frac{\p W}{\p t}+{\rm div}\left( \mathbf v W \right)
+W{\rm div}\mathbf{v}=N\left[\frac{mv^2}{\tau}-\gamma (T-T_0) \right],\notag\\
\label{dW-dt}
\EEA
where $\mathbf F_0$ is driving electric force in the homogeneous case, $- {e^2 \nabla N}/{\cal C}$ is inhomogeneity-induced correction to this force,    ${\cal C}$ is the gate-to-channel capacitance per unit length (we assume that the system is gated), and
$$W=W(N,T)= \int_0^\infty  d\epsilon  \epsilon f_F (\epsilon)$$
 is the density of energy in the moving frame (here $f_F= 1/\exp[ (\epsilon - \mu)/T +1]$ is the  Fermi function and $\mu=T\ln \left[ \exp(N/\nu T)-1\right]$), which is given by
\be
W= \left\{\begin{array}{c} \displaystyle
     \frac{N^2}{2\nu}+ \frac{\pi^2 \nu T^2}{6},\quad{\rm for}~ \nu T \ll N~  \\ \\\displaystyle
    T N, \quad{\rm for}~ \nu T \gg N.
 \end{array} \right.
\label{W-mu}
\ee
Here, we neglected the heat conductivity and viscosity of the electron liquid, setting $\tau_\text{ee}\to 0$.

Introducing heat capacitance
$$C=C(N,T)= (\p W /\p T)_{N=\text{const}}$$
 after some algebra we obtain from Eqs.~\eqref{dN-dt},\eqref{dv-dt}, and \eqref{dW-dt},  the temperature balance equation   Eq.~\eqref{dC-dt}, of the main text.

\section{Smooth constriction}
\label{app:constriction}

\begin{figure}[ptb]
\centering
\includegraphics[width=0.40\textwidth]{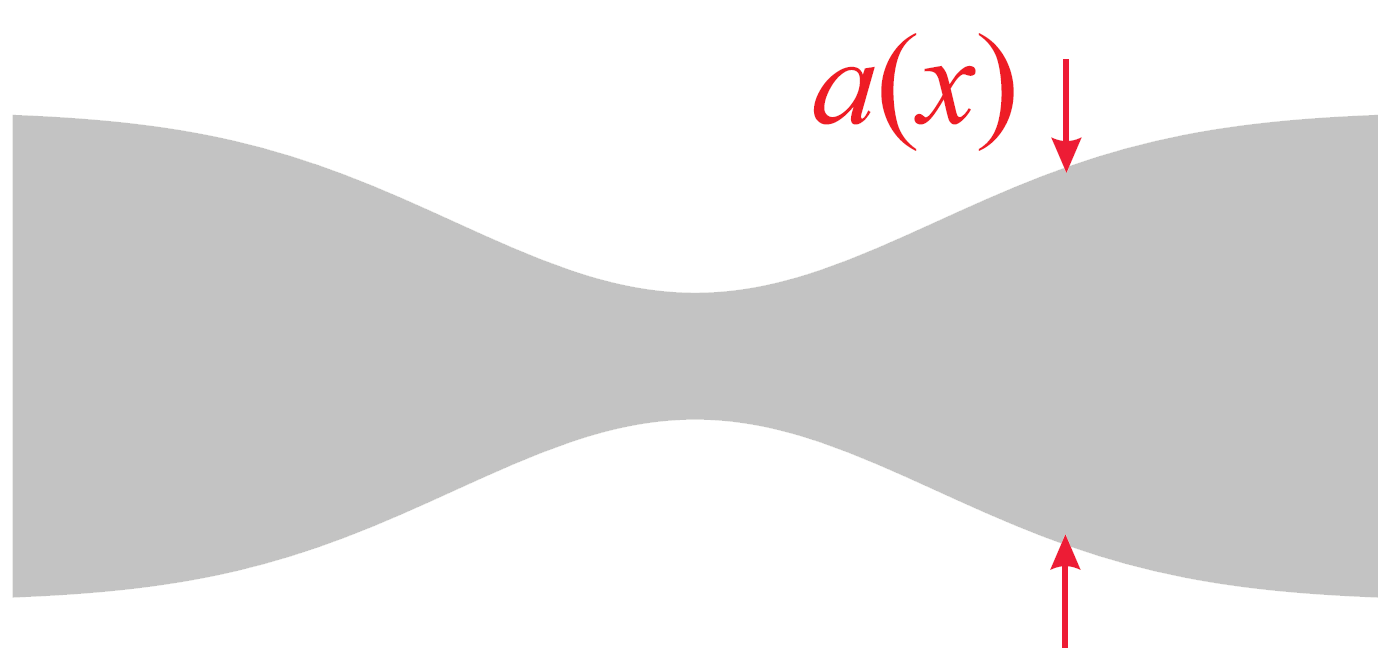}
\caption{Schematics of a smooth constriction.}
\label{fig-smooth}
\end{figure}

In this Appendix, we demonstrate that the results obtained  for the  hydrodynamic regime obtained in Sec. \ref{hydro} for a model of $\tau(x)$ inhomogeneity are in fact generic and hold also for a constriction.
Specifically, we assume that the width of the strip smoothly varies, forming a geometric constriction characterized by the local strip width $a(x)$, see Fig.~\ref{fig-smooth}. Such a constriction works effectively as an additional source of local resistance, so that even for $\tau(x)=\text{const}$ the temperature is expected to vary along the strip. This setup can be viewed as a prototype of a point contact considered in Ref.~\cite{rokni1995joule}. For incompressible electron fluid, we still assume
$N \approx N_0=\text{const}$. Then, because of the total current conservation, the current density and, hence, the drift velocity become $x$-dependent:
 \be
 v(x)=v_\infty \frac{a_\infty}{a(x)},
 \ee
where $v_\infty=F_0 \tau_\infty/m.$

%%%%%%%%%%%%%%%
%%%%%%%%%%%%%%%

Assuming that $|T(x)-T| \ll T,$
we linearize Eq.~\eqref{dC-dt-k} with respect to small variation of  $T(x)$ $v(x)$  and $\tau(x).$
In the absence of variations,  Eq.~\eqref{dC-dt-k} is satisfied because of the  identity [see Eq.~\eqref{hom}]
 \be m\frac{v_\infty}{\tau_\infty}= \frac{\gamma}{v_\infty} (T-T_0), \label{id} \ee  which relates $v_\infty$ and $T=T(x\to \pm \infty).$
In the first order, we get
\begin{align}
& \frac{C}{N_0} \left (\frac{d \delta T}{dx} + \frac{T}{v_\infty} \frac{dv}{dx}\right) 
\\
&= 
m \left ( \frac{v}{\tau} - \frac{v_\infty}{\tau_\infty}  \right )- \frac{\gamma}{v_{\infty}} \delta T - \gamma\, (T-T_0)   \left( \frac{1}{v} - \frac{1}{v_\infty}\right )
.
\notag
\end{align}
Expressing $T-T_0$ in the last term in the r.h.s. of this equation  
 with the use of Eq.~\eqref{id}, we find
that  Eq. \eqref{dT-dx-Q} is still valid with a minor modification of $G(x)$:
\be
\label{gdef2}
G(x)= \frac{m N_0}{C}\left[ \frac{v(x)}{\tau(x)}-\frac{v_\infty^2}{\tau_\infty v(x)}\right] - \frac{T}{v_\infty} \frac{dv}{dx}.
\ee
Now, Eq. (\ref{dT(x)}) is valid with $G(x)$ given by Eq. (\ref{gdef2}). As a result, for  $\tau=\text{const}$ we find
\be
\delta T_{\rm max}=   \frac{F_0 N_0}{C} \int \limits_{-\infty}^{\infty} dx \left[ \frac{a_{\infty}}{a(x)}-\frac {a(x)}{a_{\infty}} \right]
\ee
which is  analogous to  Eq.~\eqref{dTmax}.

\section{Solution of the kinetic equation in the impurity-dominated limit}
\label{app:int}

In this Appendix, we solve the linearized kinetic equation \eqref{kinur for f0} for
$\delta f$ defined in Eq. (\ref{deltaf}).
Using dimensionless variables
$$E=\epsilon/T,\quad X= x/l_{\rm in} ,   \quad Q= q l_{\rm in},$$   where $l_{\rm in}= v/\gamma$ with $v$ given by Eq.~\eqref{v-infty}, we find the following equation for the Fourier transform $\delta f_Q$ of $\delta f(x)$ introduced in Eq. \eqref{deltaf}:
\begin{equation}
\hat{L}\delta f_Q=\xi_Q S_{1}\left(E\right) +\lambda_Q S_{2}\left(E\right) .  \label{LfEq}
\end{equation}%
Here%
\begin{equation}
\hat{L} = E \p^2_E +\left( 2i Q E+E +1\right) \p_E+ 1+i Q-\frac{Q^{2} E}{\alpha}
\end{equation}
and
\begin{align}
&S_{1}\left( E \right) = e^{- E }\left( \alpha E
-i Q E - \alpha\right) ,
\\
&S_{2}\left( E \right)
= e^{- E}\left( 2 \alpha E -2 \alpha -i Q E \right).
\end{align}
The parameter  $\alpha $ characterizes the degree of overheating [see Eq.~\eqref{alpha}].
The function $\lambda_Q$ entering the right-hand side of Eq.~\eqref{LfEq} should be found self-consistently with the use of Eqs.~\eqref{lambda} and \eqref{dF}.

The requirement that the distribution function is finite both at
$E=0$ (where one of the solutions diverges logarithmically) and
at $ E =\infty $
(where one of the solutions diverges  exponentially) gives rise to discrete spectrum of the operator $\hat L$.
Eigenfunctions $f_{n}$  and  eigenvalues $L_n$ of the operator   $\hat{L}$   enumerated  by  integer index $n=0,1,2,\ldots$ read
\begin{align}
&f_{n}\left( E \right) =e^{-\frac{1}{2}
\left( 1+2i Q+Z\right) E}U\left(-n,1,Z E \right),
\\
&L _{n}=\frac{1}{2}\left( 1-Z-2nZ\right),
\end{align}
where
$$Z=\sqrt{1+4 Q\left[ i+Q\left( 1/\alpha -1 \right) \right] }$$
and $U(a,b,z)$ is the confluent hypergeometric function (polynomial in $z$ at negative integer $a$).
The functions $f_{n}\left( E \right) $  obey orthogonality condition
\[
\int_{0}^{\infty }f_{m}\left( E \right) f_{n}\left( E \right)
e^{ E \left( 1+2i Q \right) }d E =\frac{\left( n!\right) ^{2}}{Z}\delta _{mn}.
\]
Now, we can solve Eq (\ref{LfEq}) by the eigenmode expansion:%
\begin{equation}
\label{fqsol}
\delta f_Q\left( E \right) =\sum_{n=0}^{\infty }\left[ \xi_Q A_{n}^{\left(
1\right) }+\lambda_Q A_{n}^{\left( 2\right) }\right] f_{n}\left( E
\right)
\end{equation}
where for $k=1,2 $%
\begin{equation}
A_{n}^{\left( k\right) }
=\frac{1}{L_n}\frac{Z}{\left( n!\right) ^{2}}\int_{0}^{\infty }S_{k}\left(
E \right) f_{n}\left( E \right) e^{E \left( 1+2iQ
\right) }dE.
\end{equation}
We may now evaluate $\lambda_Q$. We limit ourselves with electroneutral limit $e\to\infty$, where the corresponding condition becomes $\delta N=0$. In this limit, we find
\be
\lambda_Q =- \frac{  \sum_{n=0}^{\infty }A_{n}^{\left( 1\right) }N_{n}
}{\sum_{n=0}^{\infty }A_{n}^{\left( 2\right) }N_{n}}\xi_Q .
\label{lambda-incompp}
\ee
where $$N_{n}=\int_{0}^{\infty }f_{n}\left( E \right) dE.$$

Finally, we  calculate the effective temperature of the distribution%
\begin{align}
\delta T_Q&=T\int_{0}^{\infty }\left( E -1\right)   \delta f_Q\left(
E \right) d E
\nonumber
\\
&=T\sum_{n=0}^{\infty }\left[ \xi_Q A_{n}^{\left( 1\right)
}+\lambda_Q A_{n}^{\left( 2\right) }\right] T_{n},
\label{dtQ}
\end{align}
where
$$T_{n}=\int_{0}^{\infty }\left( E -1\right)  f_{n}\left(E \right)dE.$$

It is convenient to write the final result for the temperature distribution in the form analogous to Eq.~\eqref{dT(x)}:
\be
\delta T_Q = K(Q) \xi_Q,
\label{dT-dQ}
\ee
where
 \be
 \label{KQ}
 K(Q)\!=T\;\frac{\!\sum \limits_{n=0}^{\infty}\sum \limits_{m=0}^{\infty} T_n N_m\! \left[ A_n^{(1)} A_m^{(2)}-A_n^{(2)}A_m^{(1)} \right]}{\sum \limits_{m=0}^{\infty} N_m  A_m^{(2)}}.
 \ee

The integrals determining $A_{n}^{(k)}$ and $T_n,\;N_n$ can be explicitly evaluated. For compactness, we introduce
$$\mathcal{W}_{\pm\pm}=Z\pm 2iQ\pm 1$$
to write:
\begin{align}
&A_n^{(1)}=\frac{8Z}{n!}\frac{\mathcal{W}_{+-}^{n-1}}{\mathcal{W}_{-+}^{n+2}}\left(iQ-\alpha-iQ\alpha/L_n\right),
\\
&A_n^{(2)}=\frac{8Z}{n!}\frac{\mathcal{W}_{+-}^{n-1}}{\mathcal{W}_{-+}^{n+2}}\left(1+iQ/L_n\right)\left(iQ-2\alpha\right),
\\
&T_n=-8n!\frac{\mathcal{W}_{--}^{n-1}}{\mathcal{W}_{++}^{n+2}}\left(L_n+iQ+Q^2/\alpha\right),
\\
&N_n=\frac{8n!Q^2}{\alpha}\frac{\mathcal{W}_{--}^{n-1}}{\mathcal{W}_{++}^{n+2}}.
\end{align}
Expression \eqref{KQ} is used in the main text for the analysis of the dissipation profiles
in various limiting cases.


\begin{thebibliography}{22}
\bibitem{formula}
With the Gaussian spatial profile of the inhomogeneity, $1/\tau(x)=1/\tau_\infty+ \left({1}/{\tau_0}-{1}/{\tau_\infty}\right)\exp(- {x^2}/{L^2})$, the temperature distribution can be calculated analytically for an arbitrary relation between $l_{\rm in}$ and $L$:
$
\delta T(x)=({\delta T_{\rm max}}/{2})\exp\left({ L^2}/{4 l_{\rm in}^2} -  {x}/{l_{\rm in}} \right)\!   {\rm Erfc}\!\left(\!{ L}/{2 l_{\rm in}}\! - \!{x}/{L}\!\right),
$
where $\delta T_{\rm max}=  \sqrt \pi L G(0).$
\end{thebibliography}
\end{document}